\documentclass[aps,prb,twocolumn,superscriptaddress,showpacs]{revtex4}
\usepackage{times}
\usepackage{amssymb,amsmath}
\usepackage{graphicx}
\usepackage{natbib}
\usepackage{multirow}
\usepackage{color}
\usepackage[a4paper,left=2cm,top=3cm,right=1cm,bottom=2cm]{geometry}

\begin{document}

\title{Non-exponential magnetic relaxation in magnetic nanoparticles for hyperthermia}

\author{I. Gresits}
\affiliation{Department of Non-Ionizing Radiation, National Public Health Center, Budapest, Hungary}
\affiliation{Department of Physics, Budapest University of Technology and Economics and MTA-BME Lend\"{u}let Spintronics Research Group (PROSPIN), Po. Box 91, H-1521 Budapest, Hungary}

\author{Gy. Thur\'{o}czy}
\affiliation{Department of Non-Ionizing Radiation, National Public Health Institute, Budapest, Hungary}

\author{O. S\'agi}
\affiliation{Department of Physics, Budapest University of Technology and Economics and MTA-BME Lend\"{u}let Spintronics Research Group (PROSPIN), Po. Box 91, H-1521 Budapest, Hungary}

\author{S. Kollarics}
\affiliation{Department of Physics, Budapest University of Technology and Economics and MTA-BME Lend\"{u}let Spintronics Research Group (PROSPIN), Po. Box 91, H-1521 Budapest, Hungary}

\author{G. Cs\H{o}sz}
\affiliation{Department of Physics, Budapest University of Technology and Economics and MTA-BME Lend\"{u}let Spintronics Research Group (PROSPIN), Po. Box 91, H-1521 Budapest, Hungary}

\author{B. G. M\'arkus}
\affiliation{Department of Physics, Budapest University of Technology and Economics and MTA-BME Lend\"{u}let Spintronics Research Group (PROSPIN), Po. Box 91, H-1521 Budapest, Hungary}

\author{N. M. Nemes}
\affiliation{GFMC, Unidad Asociada ICMM-CSIC "Laboratorio de Heteroestructuras con
Aplicaci\'{o}n en Espintronica", Departamento de Fisica de Materiales Universidad
Complutense de Madrid, 28040}
\affiliation{Instituto de Ciencia de Materiales de Madrid, 28049 Madrid, Spain}

\author{M. Garc\'{i}a Hern\'{a}ndez}
\affiliation{GFMC, Unidad Asociada ICMM-CSIC "Laboratorio de Heteroestructuras con
Aplicaci\'{o}n en Espintronica", Departamento de Fisica de Materiales Universidad
Complutense de Madrid, 28040}
\affiliation{Instituto de Ciencia de Materiales de Madrid, 28049 Madrid, Spain}

\author{F. Simon\email{f.simon@eik.bme.hu}}
\affiliation{Department of Physics, Budapest University of Technology and Economics and MTA-BME Lend\"{u}let Spintronics Research Group (PROSPIN), Po. Box 91, H-1521 Budapest, Hungary}
\affiliation{Laboratory of Physics of Complex Matter, \'{E}cole Polytechnique
F\'{e}d\'{e}rale de Lausanne, Lausanne CH-1015, Switzerland}

\begin{abstract}
Magnetic nanoparticle based hyperthermia emerged as a potential tool for treating malignant tumours. The efficiency of the method relies on the knowledge of magnetic properties of the samples; in particular, knowledge of the frequency dependent complex magnetic susceptibility is vital to optimize the irradiation conditions and to provide feedback for material science developments. We study the frequency-dependent magnetic susceptibility of an aqueous ferrite suspension for the first time using non-resonant and resonant radiofrequency reflectometry. We identify the optimal measurement conditions using a standard solenoid coil, which is capable of providing the complex magnetic susceptibility up to 150 MHz. The result matches those obtained from a radiofrequency resonator for a few discrete frequencies. The agreement between the two different methods validates our approach. Surprisingly, the dynamic magnetic susceptibility cannot be explained by an exponential magnetic relaxation behavior even when we consider a particle size-dependent distribution of the relaxation parameter.
\end{abstract}

\maketitle

\section*{Introduction}
Nanomagnetic hyperthermia, NMH, \cite{{PankhurstReview2003},{PankhurstHyperthermia},{Ortega_Pankhurst_Review},{KUMAR2011789},{ANGELAKERIS20171642},{GIUSTINI2010},{KRISHNAN2010},{BEIK2016205},{Spirou_2018},{delaPresa2012},{Dennis2008},{Kita2010},{Vallejo_Fernandez_2013}} is intensively studied due to its potential in tumor treatment. The prospective method involves the delivery of ferrite nanoparticles to the malignant tissue and a localized heating by an external radiofrequency (RF) magnetic field affects the surrounding tissue only. 
The key medical factors in the success of NMH\cite{{KUMAR2011789},{GIUSTINI2010},{ANGELAKERIS20171642},{PERIGO20150204}} include the affinity of tumour tissue to heating and the specificity of the targeted delivery. 

Concerning the physics and material science challenges, i) the efficiency of the heat delivery, ii) its accurate control and iii) its precise characterization are the most important ones. Concerning the latter, various solutions exists which includes modeling the exciting RF magnetic field with some knowledge about the magnetic properties of the ferrite \cite{{GARAIO2014432},{GARAIO20142511},{0957-4484-26-1-015704},{CONNORD2014},{CARREY2011}}, measurement of the delivered heat from calorimetry \cite{{GARAIO2014432},{0957-4484-26-1-015704},{WANG2013},{ESPINOSA20161002},{Bae2012}}, or determining the dissipated power by monitoring the quality factor change of a resonator in which the tissue is embedded \cite{GresitsSciRep,Gresits_2019}. 

All three challenges are related to the accurate knowledge of the frequency-dependent complex magnetic susceptibility, $\widetilde{\chi}=\chi'-i \chi''$, of the nanomagnetic ferrite material. The dissipated power per unit volume, $P$ is proportional to the value of $\chi''$ at the working frequency, $\omega$, as: $P=0.5 \mu_0 \omega \chi'' H_{\text{AC}}^2$, where $\mu_0$ is the vacuum permeability, $H_{\text{AC}}$ is the AC magnetic field strength. Although measurement of $\widetilde{\chi}(\omega)$ is a well advanced field due to e.g. the extensive filter or  transformer applications\cite{{Ahmed2018},{GHOSH2019},{Radonic2010},{Akhter2012},{Chowdary_2014},{meuche1997},{Bowler2006},{Dosoudil2013},{Kuipers2008},{Berkum2013},{delaPresa2012}}, we are not aware of any such attempts for nanomagnetic particles which are candidates for hyperthermia. 

Knowledge of $\widetilde{\chi}(\omega)$ would allow to determine the optimal working frequency, which is crucial to avoid interference due to undesired heating of nearby tissue e.g. by eddy currents \cite{PankhurstReview2003,Ortega_Pankhurst_Review,Spirou_2018}. In addition, an accurate characterization of $\widetilde{\chi}(\omega)$ can provide an important feedback to material science to improve the ferrite properties. Last but not least, measurement of $\widetilde{\chi}(\omega)$ would allow for a better theoretical description of the high frequency magnetic behavior of ferrites. Most reports suggest \cite{PankhurstReview2003,Ortega_Pankhurst_Review,Spirou_2018,IlgPRE2019} that a single relaxation time, $\tau$, governs the frequency dependence of $\widetilde{\chi}(\omega)$. The magnetic relaxation time, $\tau$, is given by to the Brown and N\'{e}el processes; these two processes describe the magnetic relaxation due to the motion of the nanomagnetic particle and the magnetization of the nanoparticle itself (while the particle is stationary). When the two processes are uncorrelated, the magnetic relaxation time is given as $1/\tau=1/\tau_{\text{B}}+1/\tau_{\text{N}}$, where ${\tau_{\text{B}}}$ and ${\tau_{\text{N}}}$ are the respective relaxation times. These two relaxation types have very different particle size and temperature dependence, which would allow for a control of the dissipation. Nevertheless, the major open questions remain, i) whether the single exponential description is valid, and ii) what the accurate frequency dependence of the magnetic susceptibility is.

Motivated by these open questions, we study the frequency dependence of $\widetilde{\chi}$ on a commercial ferrite suspension up to 150 MHz. We used two types of methods: a broadband non-resonant one with a single solenoid combined with a network analyzer and a radiofrequency resonator based approach. The latter method yields the ratio of $\chi''$ and $\chi'$ for a few discrete frequencies. The two methods give a good agreement for the frequency-dependent ratio of $\chi''/\chi'$ which validates both measurement techniques. We find that the data cannot be explained by assuming that each magnetic nanoparticle follows a magnetic relaxation with a single exponent even when the particle size distribution is taken into account. Our work not only presents a viable set of methods for the characterization of $\widetilde{\chi}$ but it provides input to the theories aimed at describing the magnetic relaxation in nanomagnetic particles and also a feedback for future material science developments.

\section{Theoretical background and methods}

The physically relevant quantity in hyperthermia is the imaginary part of the complex magnetic susceptibility, $\widetilde{\chi}$, i.e. $\chi''$ as the absorbed power is proportional to it. Although, we recently developed a method to directly determine the absorbed power during hyperthermia\cite{GresitsSciRep}, a method is desired to determine the full frequency dependence of $\widetilde{\chi}$. This would not only lead to finding the optimal irradiation frequency during hyperthermia but it could also provide an important feedback to materials development and for the understanding of the physical phenomena behind the complex susceptibility in ferrite suspensions.

The generic form of the complex magnetic susceptibility of a material reads:
\begin{gather}
\widetilde{\chi}(\omega)=\chi'(\omega)-i\chi''(\omega).
\end{gather}
Linear response theory dictates that these can be transformed to one another by a Hilbert transform\cite{Alloul2011_book,SlichterBook} as:
\begin{gather}
\chi'(\omega)=\frac{1}{\pi}\mathcal{P}\int_{-\infty}^\infty \frac{\chi''(\omega')}{\omega'-\omega} \text{d} \omega',\\
\chi''(\omega)=-\frac{1}{\pi}\mathcal{P}\int_{-\infty}^\infty \frac{\chi'(\omega')}{\omega'-\omega} \text{d} \omega',
\end{gather}
where $\mathcal{P}$ denotes the principal value integral. 

We note that we use a dimensionless volume susceptibility (invoking SI units) throughout. If a single relaxation process is present (similar to dielectric relaxation or to the Drude model of conduction, which yield $\widetilde{\epsilon}(\omega)$ and $\widetilde{\sigma}(\omega)$, respectively), the complex magnetic susceptibility takes the form:
\begin{gather}
\chi'(\omega)=\chi_0\frac{1}{1+\omega^2 \tau^2},
\label{chi1}
\end{gather}
\begin{gather}
\chi''(\omega)=\chi_0\frac{\omega \tau}{1+\omega^2 \tau^2},
\label{chi2}
\end{gather}
where $\chi_0$ is the static susceptibility.

The corresponding $\chi'$ and $\chi''$ pairs can be constructed when multiple relaxation times are present in the description of their frequency dependence. There is a general consensus\cite{{PankhurstReview2003},{Eberbeck2013},{KUMAR2011789},{BEIK2016205},{Dutz2014},{PERIGO20150204},{CARREY2011},{0957-4484-26-1-015704},{ROSENSWEIG2002},{OBEADA2013},{Dutz2013}} although experiments are yet lacking, that the single relaxation time description approximates well the frequency dependence of the magnetic nanoparticles. The frequency dependence of $\chi''$ is though to be described by the relaxation time of the nanoparticles: $1/\tau=1/\tau_{\text{N}}+1/\tau_{\text{B}}$, where the N\'eel and Brown relaxation times are related to the motion of the magnetization with respect to the particles and the motion of the particle itself, respectively.

We used a commercial sample (Ferrotec EMG 705, nominal diameter 10 nm) which contains aqueous suspensions of single domain magnetite (Fe$_3$O$_4$) nanoparticles. We verified the magnetic properties of the sample using static SQUID magnetometry; it showed the absence of a sizable magnetic hysteresis (data shown in the Supplementary Information), which proves that the material indeed contains magnetic mono-domains.

\subsection{Measurements with non-resonant circuit}
At frequencies below $\sim 5-10\,\text{MHz}$ the conventional methods of measuring the current-voltage characteristics can be used for which several commercial solutions exist. This method could e.g. yield the inductivity change for an inductor in which a ferrite sample is placed. However, above these frequencies the typical circuit size starts to become comparable to the electromagnetic radiation wavelength thus wave effects cannot be neglected. The arising complications can be conveniently handled with measurement of the $S$ parameters, i.e. the reflection or transmission for the device under test.

Obtaining $\widetilde{\chi}(\omega)$ is possible by perturbing the circuit properties of some broadband antennas or waveguides while monitoring the corresponding $S$ parameters\cite{chen2004microwave} (the reflected amplitude, $S_{11}$, and the transmitted one, $S_{21}$) with a vector network analyzer (VNA). We used two approaches: i) a droplet of the ferrite suspension on a coplanar waveguide (CPW) was measured and ii) about a 100 $\mu$l suspension was placed in a solenoid.  It is crucial in both cases to properly obtain the null measurement, i.e. to obtain the perturbation of the circuit due to the ferrite only. For the solenoid, we found that a sample holder filled with water gives no perturbation to the circuit parameters as expected. In contrast, the CPW parameters are strongly influenced by a droplet of distilled water whose quantity can be hardly controlled therefore performing the null measurement was impossible and as a result, the use of the CPW turned out to be impractical. Additional details about the VNA measurements, including details of the failure with the CPW based approach, are provided in the Supplementary Information.

\begin{figure}[htb]
\includegraphics[width=.4\textwidth]{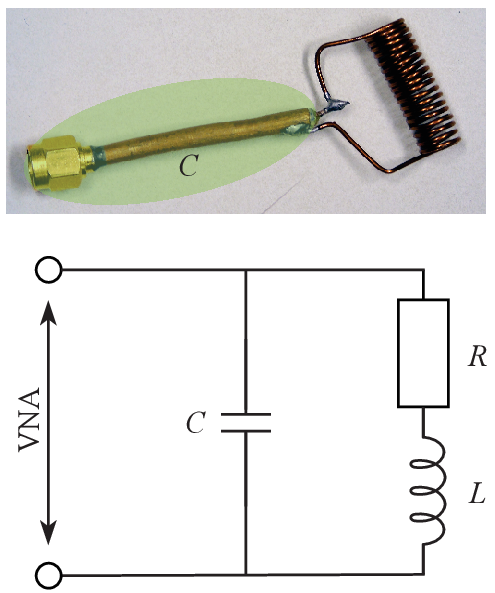}
\caption{Upper panel: photograph of the solenoid used in the non-resonant susceptibility measurements. Lower panel: the equivalent circuit model including a parasitic capacitor, $C$ due to the small coaxial cable section and the self capacitance of the inductor. $R$ has a frequency dependence due to the skin-effect.}
\caption{Upper panel: photograph of the solenoid used in the non-resonant susceptibility measurements. Lower panel: the equivalent circuit model including a parasitic capacitor, $C$ due to the small coaxial cable section and the self capacitance of the inductor. $R$ has a frequency dependence due to the skin-effect.}
\label{Solenoid_VNA}
\end{figure}

In the second approach, we used a conventional solenoid (shown in Fig. \ref{Solenoid_VNA}) made from 1 mm thick enameled copper wire, its inner diameter is 6 mm and it has a length of 23 mm with 23 turns. The coil is soldered onto a semi-rigid copper RF cable that has a male SMA connector. Fig. \ref{Solenoid_VNA}. shows the equivalent circuit which was found to well explain the reflection coefficient in the DC-150 MHz frequency range (more precisely from 100 kHz which is the lowest limit of our VNA model Rohde \& Schwarz ZNB-20). The frequency dependence of the wire re
ce due to the skin-effect was also taken into account in the analysis. The parallel capacitor arises from the parasitic self capacitance of the inductor and from the small coaxial cable section. Further details about the validation of the equivalent circuit (i.e. our fitting procedure) are provided in the Supplementary Information.

The frequency dependent complex reflection coefficient, $\Gamma$ (same as $S_{11}$ this case), and $Z$ of the studied circuit are related by \cite{PozarBook}:
\begin{equation}
\Gamma=\frac{Z-Z_0}{Z+Z_0},
\label{Gamma}
\end{equation}
where $Z_0$ is the 50 $\Omega$ wave impedance of the cables and $Z$ is the complex, frequency dependent impedance of the non-resonant circuit. It can be inverted to yield $Z$ as: $Z=Z_0\frac{1+\Gamma}{1-\Gamma}$.

The admittance for the empty solenoid reads:
\begin{equation}
\frac{1}{Z_{\text{empty}}}=\frac{1}{R(\omega)+i\omega L}+i\omega C,
\label{Zinv}
\end{equation}
The analysis yields fixed parameters for $R(\omega)$ and $C$, whereas the effect of the sample is a perturbation of the inductivity: $L\rightarrow L(1+\eta \widetilde{\chi}(\omega))$. We introduced the dimensionless filling factor parameter, $\eta$, which is proportional to the volume of the sample per the volume of the solenoid, albeit does not equal to this exactly due to the presence of stray magnetic fields near the ends of the solenoid. This parameter, $\eta$, also describes that the susceptibility can only be determined up to a linear scaling constant with this type of measurement. In principle, the absolute value of $\widetilde{\chi}(\omega)$ could be determined by calibrating the result by a static susceptibility measurement (e.g. with a SQUID magnetometer) and by extrapolating the dynamic susceptibility to DC. It is however not possible with our present setup as $\chi$ shows a strong frequency dependence down to our lowest measurement frequency of 100 kHz.

A straightforward calculation using Eq.~\eqref{Zinv} yields that $\eta \widetilde{\chi}(\omega)$ can be obtained from the measurement of the admittance in the presence of the sample, $1/Z_{\text{sample}}$ as:
\begin{equation}
\eta\widetilde{\chi}=\frac{\left(\frac{1}{Z_\text{sample}}-i\omega C\right)^{-1}-\left(\frac{1}{Z_{\text{empty}}}-i\omega C\right)^{-1}}{i\omega L}
\label{etachi}
\end{equation}

\subsection{Measurements with resonant circuit}

\begin{figure}[htp]
\begin{center}
\includegraphics[width=.45\textwidth]{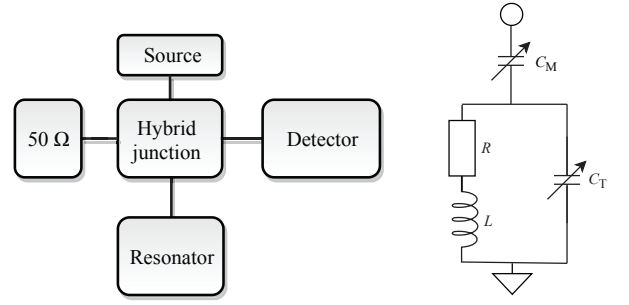}
\caption{Left: Block diagram of the resonant measurement method. Right: The schematics of the resonator circuit. It has 2 variable capacitors, the tuning ($C_\text{T}$) is used for setting the resonant frequency and the matching ($C_\text{M}$) is for setting the impedance of the resonator to 50 $\Omega$ at resonant frequency. Detailed description is in Ref. \onlinecite{GresitsSciRep}}
\label{Fig:Resonant_circuit}
\end{center}
\end{figure}

Fig. \ref{Fig:Resonant_circuit} shows the block diagram of the resonant circuit measurements which is the same as in the previous studies \cite{GresitsSciRep,Gresits_2019}. This type of measurement is based on detecting the changes in the resonator parameters, resonance frequency $\omega_0$ and quality factor, $Q$. The presence of a magnetic material induces a change in these parameters \cite{PooleBook,chen2004microwave,PozarBook} as:
\begin{equation}
\frac{\Delta \omega_0}{\omega_0}+i\Delta \left(\frac{1}{2 Q} \right) =-\eta \widetilde{\chi}
\label{resonator_perturbation}
\end{equation}
Herein, $\Delta \omega_0$ and $\Delta \left(\frac{1}{2 Q}\right)$ are changes in resonator eigenfrequency and the quality factor. The signs in Eq. \eqref{resonator_perturbation} express that in the presence of a paramagnetic material, the resonance frequency downshifts (i.e. $\Delta \omega_0<0$) and that it broadens (i.e. $\Delta \left(\frac{1}{2 Q} \right)>0$) when both $\chi'$ and $\chi''$ are positive.

The resonator measurement has a high sensitivity to minute amounts of samples \cite{chen2004microwave} however its disadvantage is that its result is limited to the resonance frequency only. Eq. \eqref{resonator_perturbation} is remarkable, as it shows that the \emph{ratio} of $\chi''$ and $\chi'$ can be directly determined at a given $\omega_0$ 
(we use throughout the approximation that $Q$ is larger than 10, thus any change to $\omega_0$ can be considered to the first order only). Namely:

\begin{equation}
\frac{\chi''}{\chi'}=-\frac{\omega_0\Delta \left(\frac{1}{2Q} \right)}{\Delta \omega_0}=-\frac{\Delta \text{HWHM}}{\Delta \omega_0}
\label{useful_ratio}
\end{equation}
where we used that the half width at half maximum, HWHM is: $\text{HWHM}=\omega_0/2Q$. The broadening of the resonator profile means that $\Delta \text{HWHM}$ is positive.

This expression provides additional microscopic information when the magnetic susceptibility can be described by a single relaxation time:
\begin{equation}
-\frac{\Delta \text{HWHM}}{\Delta \omega_0}=\frac{\chi''}{\chi'}=\omega_0 \tau
\label{simplified_useful_ratio}
\end{equation}
E.g. when a measurement at 50 MHz returns $-\frac{\Delta \text{HWHM}}{\Delta \omega_0}=1$, we then obtain directly a relaxation time of $\tau=3\,\text{ns}$. The right hand side of Eq. \eqref{simplified_useful_ratio} can be also rewritten as $\omega_0 \tau=f_0/f_{\text{c}}$, where we introduced a characteristic frequency of the particle absorption process, i.e. where $\chi''$ has its maximum. 

This description has an interesting consequence: it makes little sense to use tiny nanoparticles, i.e. to push $\tau$ to an excessively short value (or $f_\text{c}$ to a too high value). The net absorbed power reads: $P\propto f \chi''$ and when the full expression is substituted into it, we obtain $P\propto \frac{f^2}{f_{\text{c}}\left(1+\frac{f^2}{f_{\text{c}}^2} \right)}$. This function is roughly linear with $f$ below $f_{\text{c}}$ and saturates above it to a constant value. This means that an optimal irradiation frequency should be at least as large as $f_{\text{c}}$.

\section{Results and discussion}

\begin{figure}[htp]
\begin{center}
\includegraphics[width=0.45\textwidth]{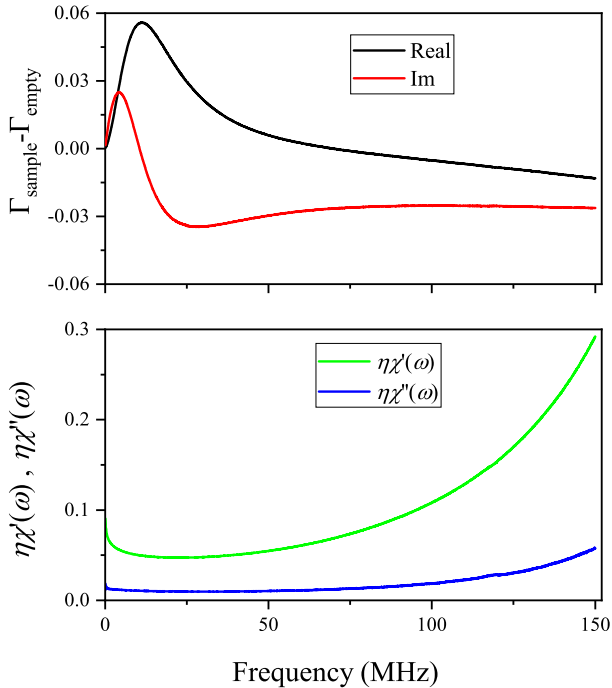}
\caption{The reflection coefficient for the solenoid with the sample inside, relative to the empty solenoid (upper panel). The real and imaginary parts of the dynamic susceptibility as obtained from the reflection coefficients according to Eqs. \eqref{Zinv} and \eqref{etachi}. Note that neither component of $\widetilde{\chi}$ follows the expected Lorentzian forms.}
\label{Fig:Gamma_chi_analysis}
\end{center}
\end{figure}

We measured the reflection coefficient for the non-resonant circuit, $\Gamma_\text{empty}$, i.e. for an empty solenoid in the 100 kHz-150 MHz range. The lower frequency limit value is set by vector network analyzer and values higher than 150 MHz are thought to be impractical due to water dielectric losses and eddy current related losses in a physiological environment \cite{{PankhurstReview2003},{PankhurstHyperthermia},{Ortega_Pankhurst_Review}}. We also measured the corresponding reflection coefficients when the ferrite suspension sample, $\Gamma_{\text{sample}}$ and only distilled water was inserted into it. The presence of the water reference does not give an appreciable change to $\Gamma$ (data shown in the Supplementary Information) as expected. The difference $\Gamma_{\text{sample}}-\Gamma_\text{empty}$ is already sizeable and is shown in Fig. \ref{Fig:Gamma_chi_analysis}. The dynamic susceptibility is obtained by first determining the empty circuit parameters (details are given in the SM) and these fixed $R$, $L$ and $C$ are used together with Eqs. \eqref{Zinv} and \eqref{etachi} to calculate $\widetilde{\chi}(\omega)$. The result is shown in the lower panel of Fig. \ref{Fig:Gamma_chi_analysis}. 

We note that the use of Eq. \eqref{etachi} eliminates $R$ and we have also checked that the result is little sensitive to about 10 \% change in the value of $L$ and $C$, therefore the result is robust and it does not depend much on the details of the measurement circuit parameters. The ratio of the two components is particularly insensitive to the parameters: $L$ cancels out formally according to Eq. \eqref{etachi} but we also verified that a 20 \% change in $C$ leaves the ratio unaffected.

\begin{figure}[htp]
\begin{center}
\includegraphics[scale=.45]{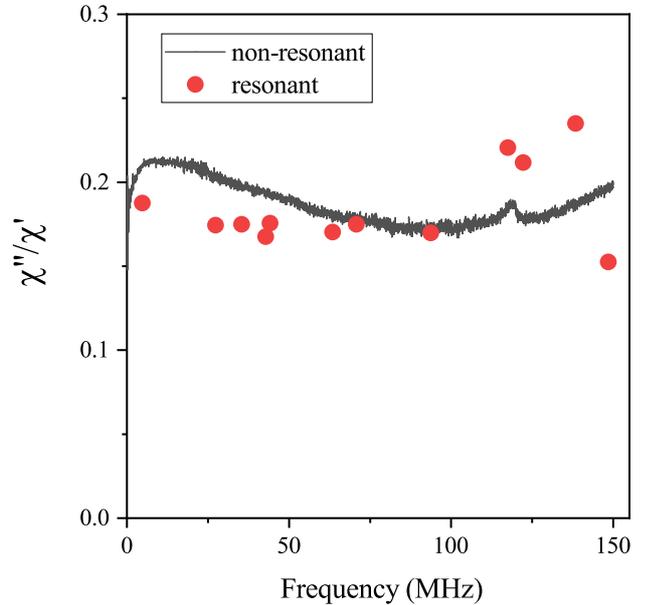}
\caption{The comparison of the ratios of $\chi$ using the non-resonant broadband and the resonator based (at some discrete frequencies) approaches.}
\label{Fig:Chi_ratios}
\end{center}
\end{figure}

Fig. \ref{Fig:Chi_ratios} shows the ratio of the two terms of the dynamics magnetic susceptibility, $\chi''(\omega)/\chi'(\omega)$, as determined by the broadband method and using the resonator based approach. The latter data is presented for a few discrete frequencies.

The two kinds of data are surprisingly close to each other, given the quite different methods as these were obtained. This in fact validates both approaches and is a strong proof that we are indeed capable of determining the complex magnetic susceptibility of the ferrofluid sample up to a high frequency.

One expects that the signal to noise performance of the resonator based approach is superior to that obtained with the non-resonant method by the quality factor of the resonator\cite{GyureGaramiJAP2019}, which is about 100. In fact, the data shows just the opposite of that and the resonator based data point show a larger scattering than the broadband approach. This indicates that the accuracy of the resonator method is limited by a systematic error, which is most probably related to the inevitable retuning of the resonator and the reproducibility of the sample placement into the resonator.

The experimentally observed dynamic susceptibility has important consequences for the practical application of hyperthermia. Given that the net absorbed power: $P\propto f \chi''$, it suggests that a reasonably high frequency, $f$, should be used for the irradiation, until other types of absorption, e.g. due to eddy currents\cite{{PankhurstReview2003},{PankhurstHyperthermia},{Ortega_Pankhurst_Review}}, limit the operation. 

We finally argue that the experimental observation cannot be explained by an exponential magnetic relaxation either due to the rotation of magnetization (the Ne\'el relaxation) or due to the rotation of the particle itself (the Brown relaxation). In principle, both relaxation processes are particle size dependent; in the nanometer particle size domain the Brown process prevails and it was calculated in Ref. \onlinecite{Ortega_Pankhurst_Review} 
that for a particle diameter of $d=10$ nm we get $\tau=300\,\text{ns}$ ($f_\text{c}=21$ MHz) for $d=11$ nm, $\tau=2\,\mu\text{s}$ ($f_\text{c}=3$ MHz), and for $d=9$ nm, $\tau=50$ ns ($f_\text{c}=125$ MHz). These frequencies would in principle explain a significant $\chi''(\omega)$ in the $1-100$ MHz range, such as we observe.

However, a simple consideration reveals from Eqs. \eqref{chi1} and \eqref{chi2} that for a single exponential magnetic relaxation for each magnetic nanoparticle, the ratio of $\chi''/\chi'$ is a \emph{straight line} as a function of the frequency, which starts from the origin with a slope depending on the distribution of the different $\tau$ parameters and particle sizes. Similarly, a single exponential relaxation would always give a monotonously decreasing $\chi'(\omega)$, irrespective of the particle size and $\tau$ distribution. Clearly, our experimental result contradicts both expectations: $\chi''/\chi'$ is not a straight line intersecting the origin and $\chi'(\omega)$ significantly increases rather than decreases above 20 MHz. We do not have a consistent explanation for this unexpected, non-exponential magnetic relaxation, which should motivate further experimental and theoretical efforts on ferrofluids. We can only speculate that a subtle interplay between the Ne\'el and Brown processes could cause this effect, whose explanation would eventually require the full solution of the equation of motion of the magnetic moment and the nanoparticles, such as it was attempted in Ref. \onlinecite{IlgPRE2019}.

\section*{Summary}
In summary, we studied the frequency-dependent dynamic magnetic susceptibility of a commercially available ferrofluid. Knowledge of this quantity is important for i) determining the optimal irradiation frequency in hyperthermia, ii) providing feedback for the material synthesis. We compare the result of two fully independent approaches, one which is based on measuring the broadband radiofrequency reflection from a solenoid and the other, which is based on using radiofrequency resonators. The two approaches give remarkably similar results for the ratio of the imaginary and real parts of the susceptibility, which validates the approach. We observe a surprisingly non-exponential magnetic relaxation for the ensemble of nanoparticles, which cannot be explained by the distribution of the magnetic relaxation time in the nanoparticles.

\section*{Acknowledgements}
The authors are grateful to G. F\"{u}l\"{o}p, P. Makk, and Sz. Csonka for the possibility of the VNA measurements and for the technical assistance. 
Jose L. Martinez is gratefully acknowledged for the contribution to the SQUID measurements. Support by the National Research, Development and Innovation Office of Hungary (NKFIH) Grant Nrs. K119442, 2017-1.2.1-NKP-2017-00001, and VKSZ-14-1-2015-0151 and by the BME Nanonotechnology FIKP grant of EMMI (BME FIKP-NAT) are acknowledged.
The authors also acknowledge the COST CA 17115 MyWAVE action.
\clearpage

\appendix
\newpage
\pagebreak
\clearpage

\section{Magnetic properties of the sample}


\begin{figure}[htp]
\begin{center}
\includegraphics[scale=0.45]{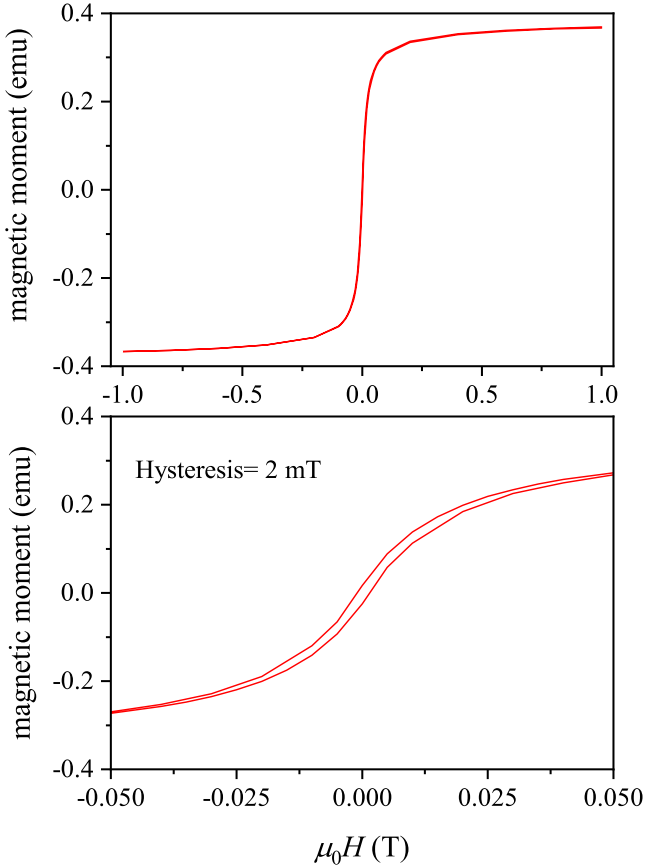}
\caption{The magnetic moment of the sample, $m$, versus the magnetic field strength, $\mu_0 H$, curve for the ferrite particle suspension. The absence of a sizeable magnetic hysteresis indicates that this is a mono\-domain sample. The estimate for the maximum hysteresis value is about 2 mT.}
\label{Fig:SQUID_M_H}
\end{center}
\end{figure}

The magnetic moment versus the magnetic field strength, $\mu_0 H$, is shown in Fig. \ref{Fig:SQUID_M_H} as measured with a SQUID magnetometer. Notably, the sample magnetism shows a saturation above 0.1 T, however it has a very small hysteresis of about 2 mT. Common hard, multidomain ferromagnetic materials, which saturate is small magnetic fields, usually display a significant hysteresis. Our observation agrees with the expected behavior of the sample, i.e. that it consists of mono-domain nanoparticle, which can easily align with the external magnetic field. 

\section{Details of the non-resonant susceptibility measurement}

\begin{figure}[htp]
\begin{center}
\includegraphics[width=0.45\textwidth]{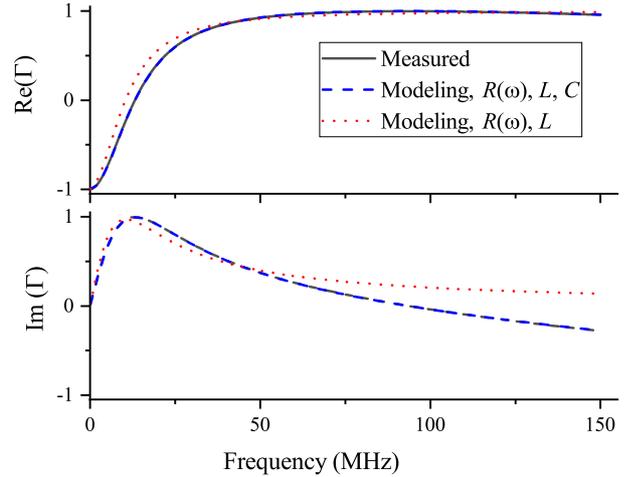}
\caption{The reflection coefficient, $\Gamma$, and its modelling with various equivalent circuit assumptions. The best fit is obtained when the equivalent circuit containing a frequency dependent wire resistance (with a small DC value) was considered in addition to an inductor and a capacitor. The absence of the capacitor does not give an appropriate fit (dotted red curve). We note that a constant wire resistance, however with an unphysically large value, gives also an appropriate fit.}
\label{Fig:Gamma_modelling}
\end{center}
\end{figure}

We discuss herein how the solenoid based broadband susceptibility measurement can be performed. We first prove that the equivalent circuit, presented in the main text, provides an accurate description. The reflectivity data is shown in Fig. \ref{Fig:Gamma_modelling}. We obtain a perfect fit (i.e. the measured and fitted curves overlap) when we consider the equivalent circuit in the main text with parameters $R_{\text{DC}}=15.5(2)\,\text{m}\Omega$, $L=0.62(1)\,\mu\text{H}$, and $C=4.65(1)\,\text{pF}$. This fit also considered the frequency dependency of the coil resistance due to the skin effect, whose DC value is $R_{\text{DC}}=13\,\text{m}\Omega$. We also performed the fit without considering the skin effect, which gave an unrealistically large $R_{\text{DC}}=150\,\text{m}\Omega$ while the fit being seemingly proper. A fit without considering a capacitor does not give a proper fit (dotted curve in the figure): its major limitation is that it cannot reproduce the zero crossing of $\Gamma$, i.e. a resonant behavior in the impedance of the circuit.
As a result, we conclude that the equivalent circuit in the main text provides a proper description of the measurement circuit and that the fitted parameters can be used to obtain the complex susceptibility of the sample, as we described in the main text.

\begin{figure}[htp]
\begin{center}
\includegraphics[width=0.47\textwidth]{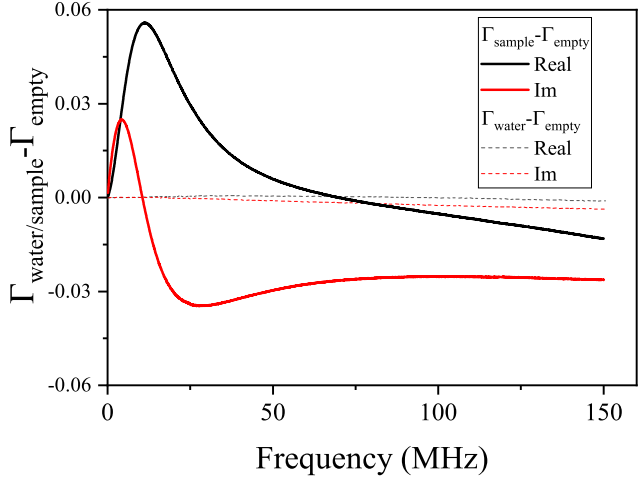}
\caption{Reflection coefficients, $\Gamma$, with respect to the empty circuit. $\Gamma_{\text{water}}$ denotes the reflection coefficient when the solenoid is filled with water in a quartz tube. Note that the sample gives rise to a significant change in the reflection below 100 MHz, whereas the presence of water (dashed lines) only slightly changes it above this frequency.}
\label{Fig:Gamma_solenoid_differences}
\end{center}
\end{figure}

Fig. \ref{Fig:Gamma_solenoid_differences}. demonstrates that the presence of the sample gives rise to a significant change in the reflection coefficient, $\Gamma$, whereas the reflection is only slightly affected by the presence of the water (maximum $\Gamma$ change is about 0.2 
\% below 150 MHz) and its effect is limited to frequencies above 150 MHz. Probably, the inevitably present stray electric fields (due to the parasitic capacitance of the solenoid) interact with the water dielectric, which results in this effect. The stray electric fields and the parasitic capacitance become significant at higher frequencies: then there is a significant voltage drop across the solenoid inductor coil, thus its windings are no longer equi-potential and an electric field emerges.

\section{Details of the susceptibility detection using a CPW}

The coplanar waveguide or CPW is a planar RF and microwave transmission line whose impedance is 50 $\Omega$ at a wide frequency range\cite{PozarBook,chen2004microwave}. The CPW can be thought of as a halved coaxial cable which makes the otherwise buried electric and magnetic fields available to study material parameters, essentially as a small piece of an irradiating antenna \cite{PozarBook,chen2004microwave}. However as we show below, the inevitable simultaneous presence of the electric and magnetic field hinders a meaningful analysis.

\begin{figure}[htp]
\begin{center}
\includegraphics[width=0.5\textwidth]{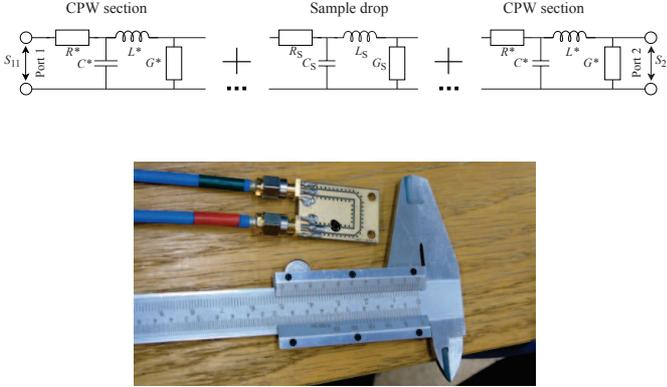}
\caption{Upper panel: The equivalent circuit of a CPW section with a sample on the top. $L_{\text{S}}$, $R_{\text{S}}$, $C_{\text{S}}$, and $G_{\text{S}}$ are effective inductance, series resistance, capacitance, and shunt conductance of the small CPW section which contains the sample, respectively. $L^{\ast}$, $R^{\ast}$, $C^{\ast}$, and $G^{\ast}$ are corresponding distributed circuit parameters which are normalized to unit length. In an ideal waveguide $R^{\ast}=0$ and $G^{\ast}=0$, and also $Z_0=\sqrt{L^{\ast}/C^{\ast}}$. Lower panel: Photo of CPW with a droplet of the sample. Port 1 is labeled with green and Port 2 is with red tape.}
\label{Fig:CPW_photo}
\end{center}
\end{figure}

We show the U shaped CPW section use in our experiments along with the equivalent circuit of the CPW in Fig. \ref{Fig:CPW_photo}. As this device has two ports, one can measure the frequency dependent complex reflection ($S_{11}$) and transmission ($S_{21}$) coefficients simultaneously with a VNA. We placed the sample droplet on the top of the gap section of the CPW between the central conductor and the grounding side plate, where the RF magnetic field component is the strongest. The presence of the sample influences all parameters for the waveguide section where it is placed: the inductance $L_{\text{s}}$, capacitance $C_{\text{s}}$, the series resistance $R_{\text{s}}$, and the shunt inductance $G_{\text{s}}$. All 4 parameters are extensive, i.e. these depend on the quantity of the sample and one can express the inductivity as $L_{\text{s}}=L_0(1+\eta\widetilde{\chi})$ where $L_0$ is the inductivity of the CPW section which is affected by the sample, $\eta$ is the relevant filing factor that is dimensionless and $\widetilde{\chi}$ is the complex magnetic susceptibility. 

The $S$ parameters for such a device read\cite{GoodFMRReview}:

\begin{equation}
S_{11,\text{sample}}=\frac{R_{\text{s}}+i\omega L_{\text{s}}+\frac{Z_0}{1+Z_0\left(G_{\text{s}}+i\omega C_{\text{s}}\right)}-Z_0}{R_{\text{s}}+i\omega L_{\text{s}}+\frac{Z_0}{1+Z_0\left(G_{\text{s}}+i\omega C_{\text{s}}\right)}+Z_0}
\label{CPW_S11}
\end{equation}
and
\begin{equation}
S_{21,\text{sample}}=\frac{2\frac{Z_0}{1+Z_0\left(G_{\text{s}}+i\omega C\right)}}{R_{\text{s}}+i\omega L_{\text{s}}+\frac{Z_0}{1+Z_0\left(G_{\text{s}}+i\omega C_{\text{s}}\right)}+Z_0}
\label{CPW_S21}
\end{equation}

We calibrated the system that without sample (empty case) so that the VNA shows 0 for $S_{11}$ and real 1 to $S_{21}$ on the entire frequency range. During the calibration, port 1 of the VNA was connected to the CPW and we assembled and disassembled the necessary calibrating elements (OPEN, SHORT, MATCH) onto port 2 and the second end of the CPW. Therefore the VNA reference plane was this end of the CPW. The calibration could be achieved down to $\Gamma<5\cdot 10^{-4}$ (not shown). 

\begin{figure}[htp]
\begin{center}
\includegraphics[width=0.5\textwidth]{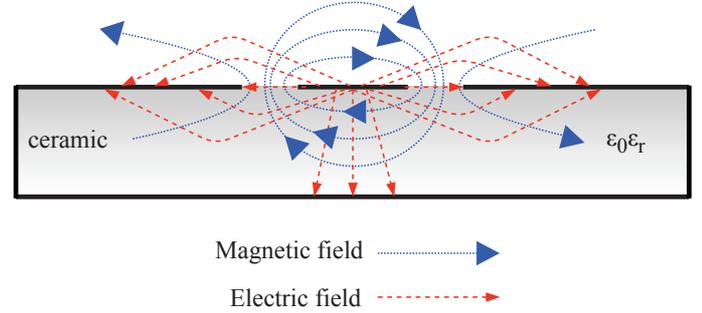}
\caption{Cross section of a coplanar waveguide showing the electric and magnetic field. Note that magnetic field is present around the central conductor and that there is a significant electric field in the two gaps between the central conductor and the neighboring ground plates.}
\label{Fig:CPW_crossection}
\end{center}
\end{figure}

We first measured the reflection transmission coefficient change under the influence of a small distilled water droplet with approximately the same size as that of the sample. We observe a $\Gamma$ change up to about 5 \% (maximum value at 150 MHz, data not shown) for both coefficients, which is a sizeable value. We note that the solenoid investigation, which we discussed in the main paper, gave a change in $\Gamma$ for the influence of water of about 0.2 \%. Clearly, this larger sensitivity of the measurement for water is due to the electric field which is significant for the CPW and is much smaller for the solenoid. 

It is even more intriguing that the effect of the sample is primarily to shift the real parts of both $S_{11}$ and $S_{21}$ by the same amount even at DC, while leaving the imaginary components unchanged (data not shown). For our typical droplet size, such as this shown in Fig. \ref{Fig:CPW_photo}, this amount is $\Delta \Gamma \approx 0.04$. Rewriting Eqs. \eqref{CPW_S11} and \eqref{CPW_S21} in the zero frequency limit, yields:

\begin{gather}
S_{11,\text{sample, DC}}=\frac{R_{\text{s}}+\frac{Z_0}{1+Z_0 G_{\text{s}}}-Z_0}{R_{\text{s}}+\frac{Z_0}{1+Z_0 G_{\text{s}}}+Z_0}, \\
S_{21,\text{sample, DC}}=\frac{2\frac{Z_0}{1+Z_0 G_{\text{s}}}}{R+\frac{Z_0}{1+Z_0 G_{\text{s}}}+Z_0}.
\label{CPW_S_low}
\end{gather}

We find that in the reasonable limit of $R_{\text{sample}}\lesssim Z_0$, the influence of $G_{\text{s}}$ dominates and that the experimental finding implies the presence of a significant shunt conductance due to the sample. We speculate that this may be due to the presence of excess OH$^-$ ions in the ferrofluid (the Ferrotec EMG 705 has a pH of 8-9), which conduct the electric current. Again, this effect is the result of the finite electric field across the gap of the CPW, where we place the sample.

The two effects, the presence of a significant capacitance due to water and a shunt inductance due to the conductivity of the ferrofluid, occur simultaneously when using a CPW for the measurement. In fact, the effect of these factors dominate the reflection/transmission. This means that determining the magnetic susceptibility for a case when a finite electric field is present, proves to be impractical.

\section{Additional details on the theory of resonators}

\begin{figure}[htp]
\begin{center}
\includegraphics[scale=0.45]{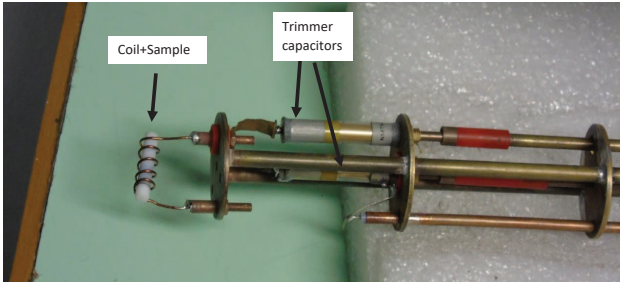}
\caption{Photograph of the radiofrequency resonator. The sample, measurement coil and the two trimmer capacitors are indicated.}
\label{Fig:SM_NMR_resonator}
\end{center}
\end{figure}

Fig. \ref{Fig:SM_NMR_resonator}. shows a photograph of the radiofrequency resonator circuit which was used in the studies. Note the presence of the two trimmer capacitors, which act as frequency tuning and impedance matching elements.

The following equation was used in the main text to determine the relation between resonator parameters and the material properties:

\begin{equation}
\frac{\Delta \omega_0}{\omega_0}+i\Delta \left(\frac{1}{2 Q} \right) =-\eta \widetilde{\chi}
\label{SM_resonator_perturbation}
\end{equation}

We note that the - sign before the imaginary term on the left hand side varies depending on the definition of the sign in the complex response function $\widetilde{\chi}$. We use the convention of Ref. \onlinecite{chen2004microwave} where $\widetilde{\chi}=\chi'-i \chi''$ which results in the + sign in Eq. \eqref{SM_resonator_perturbation}. 

The factor 2 in Eq. \eqref{resonator_perturbation} may seem disturbing but it is the direct consequence of the $Q$ factor definition: $Q=\text{FWHM}/\omega_0$, where FWHM is the full width at half maximum of the resonance curve (in angular frequency units). Thus $1/2Q=\text{HWHM}/\omega_0$, where HWHM is the half width at half maximum of the resonance curve. We also recognize that a Lorentzian shaped resonator profile can be expressed as $\frac{1}{(\omega-\omega_0)^2+1/\tau^2}$, where $\tau$ is the time constant of the resonator and $\tau=2Q/\omega_0$. This also means that $\text{HWHM}=1/\tau$. 

This allows to express the above equation in a more compact way by introducing the complex angular frequency of the resonator:
\begin{equation}
\widetilde{\omega}=\omega_0+i\frac{\omega_0}{2Q}=\omega_0+i\frac{1}{\tau}.
\label{complex_omega}
\end{equation}
It is interesting to note that the complex Lorentzian lineshape profile is proportional to $1/i\widetilde{\omega}$. It then follows from Eq. \eqref{complex_omega} that Eq. \eqref{SM_resonator_perturbation} can be expressed as:
\begin{equation}
\frac{\Delta \widetilde{\omega}}{\omega_0}=-\eta \widetilde{\chi}
\end{equation}
where $\Delta \widetilde{\omega}$ is the shift (or change) of (the complex) $\widetilde{\omega}$.

\begin{figure}[htp]
\begin{center}
\includegraphics[scale=.5]{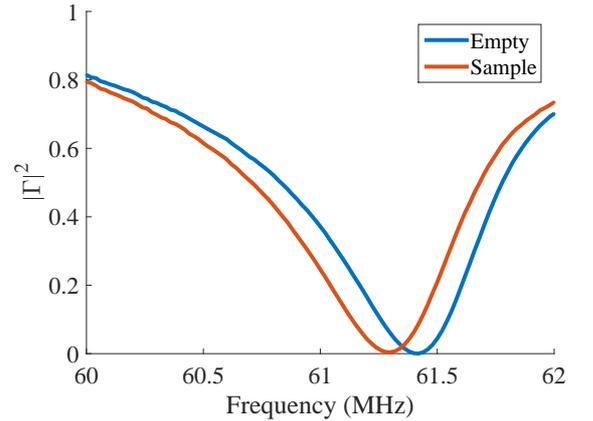}
\caption{The reflection curves at the resonant method. The shift of the resonance frequency due to the sample (red) is clearly visible.}
\label{Fig:Resonant_shift}
\end{center}
\end{figure}

Fig. \ref{Fig:Resonant_shift} shows the changes in the reflection curves at the resonant method.

\end{document}